\documentclass[preprint,showpacs,preprintnumbers,amsmath,amssymb]{revtex4}
\usepackage{graphicx}
\usepackage{dcolumn}
\usepackage{bm}
\usepackage{natbib}

\begin{document}

\preprint{}

\title{Dynamic Stability and Thermodynamic Characterization in an Enzymatic Reaction at the Single Molecule Level}

\author{Mois\'es Santill\'{a}n}
\email{msantillan@cinvestav.mx}
\affiliation{Centro de Investigaci\'{o}n y Estudios Avanzados del IPN, Unidad Monterrey, Parque de Investigaci\'{o}n e Innovaci\'{o}n Tecnol\'{o}gica, 66600 Apodaca NL, M\'{E}XICO}

\begin{abstract}
In this work we study, at the single molecular level, the thermodynamic and dynamic characteristics of an enzymatic reaction comprising a rate limiting step. We investigate how the stability of the enzyme-state stationary probability distribution, the reaction velocity, and its efficiency of energy conversion depend on the system parameters. We employ in this study a recently introduced formalism for performing a multiscale thermodynamic analysis in continuous-time discrete-state stochastic systems.

\vspace{5mm}

\end{abstract}


\maketitle

\section{Introduction}

The vast majority of processes taking place inside a cell consist of, or at least involve to some extent, chemical reactions. Moreover, most, if not all of the chemical reactions taking place inside a cell are catalyzed by enzymes. Therefore, a profound understanding of enzymes' performance is necessary to better comprehend the processes of life.

Homeostasis, understood as the coordinated physiological processes which maintain most of the steady states in an organism, is regarded as a landmark concept in biology \citep{Holmes86}. It can be found, to some extent, in all living beings, and allows them to perform in optimal conditions despite ever-changing surroundings and inputs. From a dynamical standpoint, homeostasis implies the existence of a stable steady state \citep{Kitano04,Kitano07,Lesne08}. Thus, the quality of homeostasis can be measured by the volume of the steady-state basin of attraction in phase space and/or the relaxation time with which the system returns to the steady state after a perturbation. Having a large basin of attraction is important because it allows the system to come back to the steady state even in the face of large deviations. On the other hand, a rapid relaxation time permits the system to quickly recover an optimal state after it is perturbed.

Recent studies on finite-time thermodynamic engines and heat pumps have shown that their stability and their thermodynamic performance are often governed by the same parameters. It has been observed that the system stability usually weakens as its thermodynamic properties improve \citep{SantillanEtAl01, GuzmanEtAl05, ChimalEtAl06, PaezEtAl06, Huang07, ChimalEtAl07, NieEtAl08, NieEtAl08b, HuangSun08, HuangSun08b, HuangSun08c, HuangEtAl09, Huang09}. Consequently, these parameters need to be tuned to achieve an optimal trade-off between favorable thermodynamic and dynamic properties. Similar studies on the stretch-reflex regulatory pathway and on a simple Brownian motor have confirmed these findings \citep{PaezSantillan08, SantillanMackey08}, in agreement with the notion that good design principles are usually shared by both artificial and biological systems \citep{Kitano04}. If these results are of general applicability to a wide range of intracellular energy-converting processes, it would mean that the maintenance of the cell homeostatic state entails an expenditure of energy, which has to be taken into consideration to understand how organisms adapt to a constantly changing environment.

In the present work we study, at the single molecular level, the thermodynamic and dynamic characteristics of an enzymatic reaction. For the sake of simplicity we advocate to enzymatic reactions comprising a rate limiting step. We investigate how the stability of the enzyme-state stationary probability distribution, the reaction velocity, and its efficiency of energy conversion depend on the system parameters. We employ in this study a recently introduced formalism for performing a multiscale thermodynamic analysis in continuous-time discrete-state stochastic systems. 

\section{Modeling Enzymatic Reactions}

\begin{figure}[htb]
\includegraphics[width=4in]{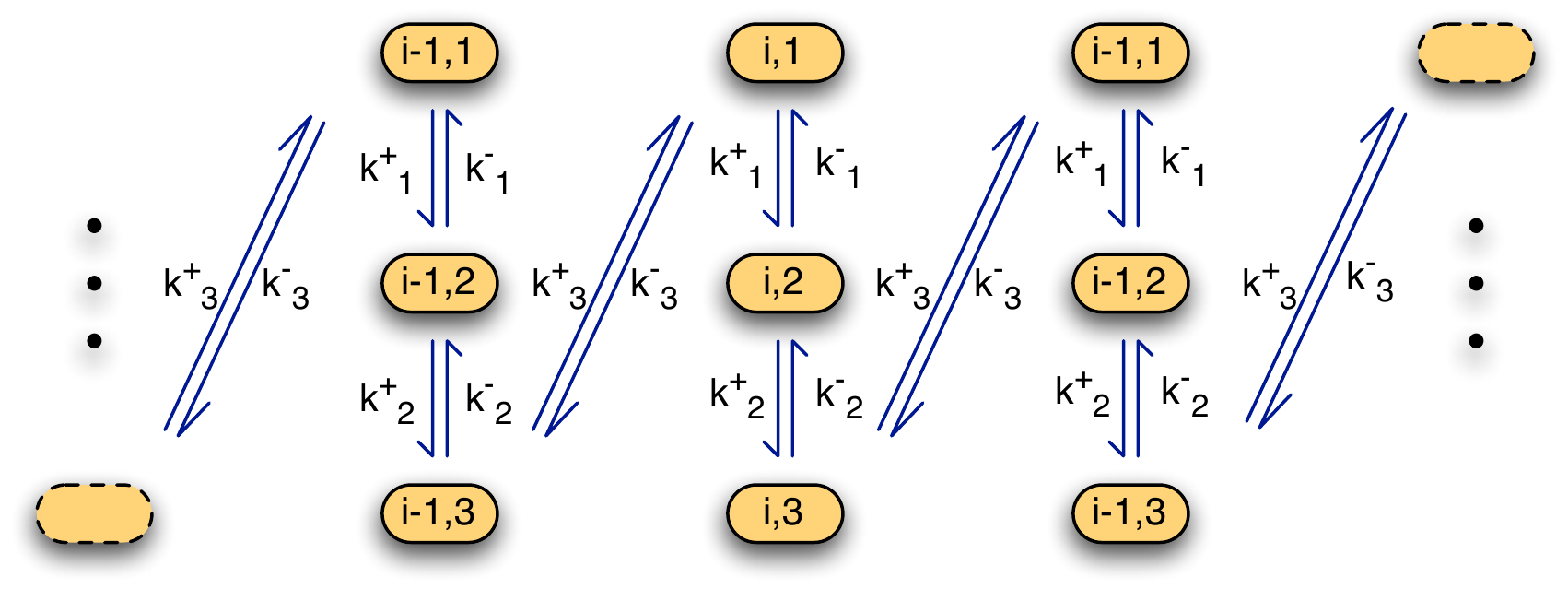}
\caption{Schematic representation of the various states available for an enzyme molecule and the transitions rates among them, while the enzyme is catalyzing the synthesis of molecules $P$. The states are denoted as $(i,j)$, where $j=1,2,3$ respectively correspond to states $E_P$, $E+P$, and $E_S$, while the index $i$ denotes the $i$ th enzyme cycle: that in which the $i$ th product molecule is synthesized and released.}
\label{Scheme}
\end{figure}

A simple but comprehensive enough model for an enzymatic reaction consists in picturing the enzyme as undergoing the following series of chemical reactions \citep{Qian:2002fk}: first, a free enzyme $E$ binds the substrate $S$; then, the bound substrate is converted into the product $P$ to form the enzyme-product complex $E_P$; and finally, the product is released leaving the enzyme free to catalyze another reaction. This process can be summarized as follows using the conventional notation for chemical reactions:
\[
E + S \rightleftharpoons E_S \rightleftharpoons E_P \rightleftharpoons E + P.
\]
Of all these reactions, the conversion of the substrate $S$ into the product $P$ ($E_S \rightleftharpoons E_P$) is in many cases the rate limiting process. Taking this into account and assuming that the substrate ($[S]$) and product ($[P]$) concentrations remain constant along the catalytic reaction, we can visualize a single enzyme as going through a series of transitions that change the enzyme state cyclically during the catalytic process; see Figure \ref{Scheme}. There, the enzyme state is represented as $(i,j)$, where index $i$ denotes the $i$ th cycle---that in which the $i$ th product molecule is synthesized---while $j=1,2,3$ respectively correspond to states $E_P$, $E+P$, and $E_S$. The assumption that $[S]$ and $[P]$ remain constant allows us to regard $k^+_j$ and $k^-_j$ ($j=1,2,3$) as pseudo first order reaction rates. Finally, the assumption that $E_S \rightleftharpoons E_P$ is the rate limiting process implies that vertical transitions in the scheme of Figure \ref{Scheme} are much faster processes than those involving changes in index $i$.

\section{Probabilistic Description and Time-Scale Separation}

Let us introduce a probabilistic description for an enzyme reaction at the single molecule level. And let us follow as well the approach in \citep{Zeron:2010uq, SantillanQian2011} to simplify the model, taking advantage of the assumed separation of time scales. Let $P(i,j;t)$ denote the probability that the enzyme is in state $(i,j)$ at time $t$. From the scheme in Figure \ref{Scheme}, the chemical master equation for $P(i,j;t)$ consists of the following set of coupled differential equations:
\begin{eqnarray}
\frac{d P(i,1;t)}{dt} & = & k^+_3 P(i-1,3;t) + k^-_1 P(i,2;t) - (k^-_3 + k^+_1) P(i,1;t),
\label{cme1} \\
\frac{d P(i,2;t)}{dt} & = & k^+_1 P(i,1;t) + k^-_2 P(i,3;t) - (k^-_1 + k^+_2) P(i,2;t),
\label{cme2} \\
\frac{d P(i,3;t)}{dt} & = & k^+_2 P(i,2;t) + k^-_3 P(i+1,1;t) - (k^-_2 + k^+_3) P(i,3;t).
\label{cme3} 
\end{eqnarray}

The probability that the enzyme is in state $(i,\cdot)$ at time $t$ is given by
\begin{equation}
P(i;t) = \sum_{j=1}^3 P(i,j;t).
\label{pslow}
\end{equation}
On the other hand, it follows from the definition of conditional probability that
\begin{equation}
P(i,j;t) = P(j|i;t) P(i;t).
\label{condprob}
\end{equation}
Add Equations (\ref{cme1})-(\ref{cme3}) and use (\ref{pslow}) and (\ref{condprob}) to obtain
\begin{eqnarray}
\frac{dP(i;t)}{dt} & = & \left( k^+_3 P(3|i-1;t) \right) P(i-1;t) + \left( k^-_3 P(1|i+1;t) \right) P(i+1;t) 
\nonumber \\
& - & \left( k^+_3 P(3|i;t) + k^3_- P(1|i; t) \right) P(i;t).
\label{cmered1}
\end{eqnarray}
Differentiate (\ref{condprob}) and assume a time-scale separation so that the transitions between states $(i-1,1)$ and $(i,3)$ are much slower than all the other (that is, they are the rate limiting steps along the reaction chain). Then,
\begin{eqnarray}
\frac{d P(1|i;t)}{dt} & = & k^-_1 P(2|i;t) + k^+_1 P(1|i;t),
\label{cmecondprob1}\\
\frac{d P(2|i;t)}{dt} & = & k^+_1 P(1|i;t) - k^-_1 P(2|i;t) + k^-_2 P(3|i;t) - k^+_2 P(2|i;t),
\label{cmecondprob2}\\
\frac{d P(3|i;t)}{dt} & = & k^+_2 P(2|i;t) - k^-_2 P(3|i;t).
\label{cmecondprob3}
\end{eqnarray}
If we invoke once more the separation of time scales to assume that the fast dynamics rapidly reach an equilibrium distribution while slow dynamics have not changed noticeably: $d P(j|i;t) / dt \approx 0$, $j=1,2,3$, we have that
\begin{eqnarray}
P(1|i;t) & \simeq & P^*(1|i), 
\nonumber \\
P(2|i;t) & \simeq & P^*(2|i) = K_1 P^*(1|i), 
\nonumber \\
P(2|i;t) & \simeq & P^*(3|i) = K_2 P^*(2|i) = K_1 K_2 P^*(1|i),
\nonumber
\end{eqnarray}
where $K_j = k^+_j / k^-_j$, $j=1,2,3$. Finally, the normalization condition ($\sum_{j=1}^3 P^*(j|i) = 1$) implies that
\begin{eqnarray}
P^*(1|i) & = & \frac{1}{1+K_1+K_1K_2},
\label{p1cond} \\
P^*(2|i) & = & \frac{K_1}{1+K_1+K_1K_2},
\label{p2cond} \\
P^*(3|i) & = & \frac{K_1 K_2}{1+K_1+K_1K_2}.
\label{p3cond} 
\end{eqnarray}
Notice that $P^*(j|i)$ ($j=1,2,3$) are all independent of $i$. Furthermore, it is straightforward to prove that this stationary distribution satisfies the following relations:
\[
k^+_1 P^*(1|i) = k^-_1 P^*(2|i), \quad \text{and} \quad k^+_2 P^*(2|i) = k^-_2 P^*(3|i).
\]
This last result implies that the fast dynamics are in equilibrium only if each one of the underlying chemical reactions is in equilibrium itself. Finally, substitution of (\ref{p1cond})-(\ref{p3cond}) into (\ref{cmered1}) allows us to conclude that, when time-scale separation is possible and the enzyme states are grouped as sketched in Figure \ref{Scheme}, the system dynamics is that of a biased one-dimensional random walk:
\begin{equation}
\frac{dP(x;t)}{dt} = k^+ P(i-1;t) + k^- P(i+1;t) - \left( k^+ + k^- \right) P(i;t),
\label{cmered}
\end{equation}
where
\begin{equation}
k^+ = \gamma \frac{K_1K_2K_3}{1+K_1+K_1K_2}, \quad \text{and} \quad k^- = \gamma \frac{1}{1+K_1+K_1K_2},
\label{kpkm}
\end{equation}
while $\gamma = k^-_3$ and $K_3 = k^+_3 / k^-_3$.

From (\ref{cmered}), the slow-dynamics stationary distribution obeys 
\[
P^*(i) = P^* \equiv \text{constant}.
\]
Observe that this distribution does not fulfill detailed balance since $k^+ P^*(i-1) - k^- P^*(i) \neq 0$, unless $k^* = k^-$.

\section{Thermodynamic State Variables and Relaxation to the Stationary State}

Following \citep{Esposito:2007uq,Ge:2010kx, SantillanQian2011}, the enzyme internal energy, entropy, and Helmholtz free energy can be respectively defined as follows:
\begin{eqnarray}
U & = & - k_B T \sum_{i,j} P(i,j;t) \log P^*(i,j),
\nonumber \\
S & = & - k_B \sum_{i,j} P(i,j;t) \log P(i,j;t),
\nonumber \\
F & = & U - TS = k_B T \sum_{i,j} P(i,j;t) \log \frac{P(i,j;t)}{P^*(i,j)}.
\nonumber 
\end{eqnarray}
By substituting (\ref{condprob}) into the above equations they can be rewritten as
\begin{eqnarray}
U & = & - k_B T \sum_{i} P(i;t) \log P^*(i) - k_B T \sum_i P(i;t) \sum_j P(j|i;t) \log P^*(j|i),
\label{energy} \\
S & = & - k_B \sum_{i} P(i;t) \log P(i;t) - k_B \sum_i P(i;t) \sum_j P(j|i;t) \log P(j|i;t),
\label{entropy} \\
F & = & k_B T \sum_{i} P(i;t) \log \frac{P(i;t)}{P^*(i)} + k_B T \sum_i P(i;t) \sum_j P(j|i;t) \log \frac{P(j|i;t)}{P^*(j|i)}.
\label{freeenergy} 
\end{eqnarray}
Observe that this way of writing the thermodynamic state variables renders a natural separation of contributions from the fast and slow dynamics.

The first term in the right hand side of Equation (\ref{freeenergy}) is nothing else but the Kullback-Leibler divergence between distributions $P(i;t)$ and $P^*(i)$ and so it is positive defined and only equals zero when the two distributions are identical. Similarly, the sum over $j$ in the second term is the Kullback-Leibler divergence between $P(j|i;t)$ and $P^*(j|i)$, it is positive defined, and only equals zero when $P(j|i;t)=P^*(j|i)$. From these considerations, the value of $F$ can be used as an indicator of how far the system is from the stationary distribution.

After differentiating Equation (\ref{freeenergy}) and imposing the separation of time scales we obtain
\begin{equation}
\frac{dF}{dt} = Q_{kh} - T \sigma,
\label{dfdt}
\end{equation}
where $Q_{hk}$ is known as the housekeeping heat and is given by
\begin{eqnarray}
Q_{hk} & = & k_B T \sum_i \left( P(i;t) k^+ - P(i+1;t) k^- \right) \log \frac{P^*(i) k^+}{P^*(i+1) k^-}
\nonumber \\
& + & k_B T \sum_i P(i;t) \sum_{j=1}^{2} \left( P(j|i;t) k^+_j - P(j+1|i;t) k^-_j \right) \log \frac{P^*(j|i) k^+_j}{P^*(j+1|i) k^-_j},
\label{qhk}
\end{eqnarray}
while $\sigma$ is the entropy production rate:
\begin{eqnarray}
\sigma & = & k_B \sum_i \left( P(i;t) k^+ - P(i+1;t) k^- \right) \log \frac{P(i;t) k^+}{P(i+1;t) k^-}
\nonumber \\
& + & k_B \sum_i P(i;t) \sum_{j=1}^{2} \left( P(j|i;t) k^+_j - P(j+1|i;t) k^-_j \right) \log \frac{P(j|i;t) k^+_j}{P(j+1|i;t) k^-_j}.
\label{sigma}
\end{eqnarray}
Observe that both $Q_{hk}$ and $\sigma$ have contributions from the slow (first term on the right hand side) and fast dynamics (second term). However, the fast dynamics contribution to $Q_{hk}$ vanishes because $P^*(j|i)$ complies with detailed balance, and so
\begin{equation}
Q_{hk} = k_B T \sum_i \left( P(i;t) k^+ - P(i+1;t) k^- \right) \log \frac{P^*(i) k^+}{P^*(i+1) k^-}.
\label{qhknew}
\end{equation}
This result is in complete agreement with the interpretation of $Q_{hk}$ as the energy that has to be pumped into the system to drive the stationary state out from equilibrium (detailed balance).

In the present formalism, the enzyme molecule is implicitly assumed to be in equilibrium with a thermal bath. Concomitantly, the proper thermodynamic description is that of the Helmholtz free energy. As seen in Equation (\ref{freeenergy}), $F\geq0$ and it only equals zero at the steady state. In other words, the value of $F$ can be used as a measure of how distant the system state is from the stationary one, as we have previously asserted. Furthermore, it is not hard to prove from (\ref{dfdt}), (\ref{sigma}), and (\ref{qhknew}) that $dF /dt \leq 0$, and that $dF /dt = 0$ only when $P(i;t) = P^*(i)$ and $P(j|i;t) = P^*(j|i)$. Therefore, $dF/dt$ can be interpreted as the rate of relaxation to the stationary distribution.

We can see from (\ref{dfdt}), (\ref{sigma}), and (\ref{qhknew}) that $dF /dt$ can be decomposed into contributions from the slow and fast dymamics:
\begin{equation}
\frac{dF}{dt} = \dot{F}_{\text{slow}} + \dot{F}_{\text{fast}},
\label{dotfglobal}
\end{equation}
where
\begin{equation}
\dot{F}_{\text{slow}} = k_B \sum_i \left( P(i;t) k^+ - P(i+1;t) k^- \right) \log \frac{P^*(i) P(j+1|i;t)}{P^*(i+1) P(j|i;t)},
\label{dotfslow}
\end{equation}
and
\begin{equation}
\dot{F}_{\text{fast}} = k_B \sum_i P(i;t) \sum_{j=1}^{2} \left( P(j|i;t) k^+_j - P(j+1|i;t) k^-_j \right) \log \frac{P(j|i;t) k^+_j}{P(j+1|i;t) k^-_j}.
\label{dotffast}
\end{equation}

The relaxation of the fast dynamics subspace to the corresponding quasi-stationary state---given by the slow-dynamics probability distribution---is determined by $\dot{F}_{\text{fast}}$. According to the assumed separation of time scales, fast dynamics relaxation takes place without any noticeable change in the slow dynamics probability distribution ($P(i;t)$) and, in the long run, the slow dynamics are the ones that govern the system relaxation to the steady state. That is, the system relaxation rate is well approximated by $\dot{F}_{\text{slow}}$.

Under the assumption that the probability distribution $P(i;t)$ is slightly different from the stationary distribution $P^*$: $P(i;t) = P^* + \epsilon(i;t)$, with $\epsilon(i;t) \ll 1$. Then, it is straightforward to see from (\ref{dotfslow}) that, in the neighborhood of the stationary distribution, the system relaxation rate (here denoted by $\xi$) is proportional to---see equation (\ref{kpkm}):
\begin{equation}
\xi \propto k^+ - k^- = \gamma \frac{K_1K_2K_3-1}{1+K_1+K_1K_2}.
\label{relrate}
\end{equation}

\begin{figure}[htb]
\includegraphics[width=3in]{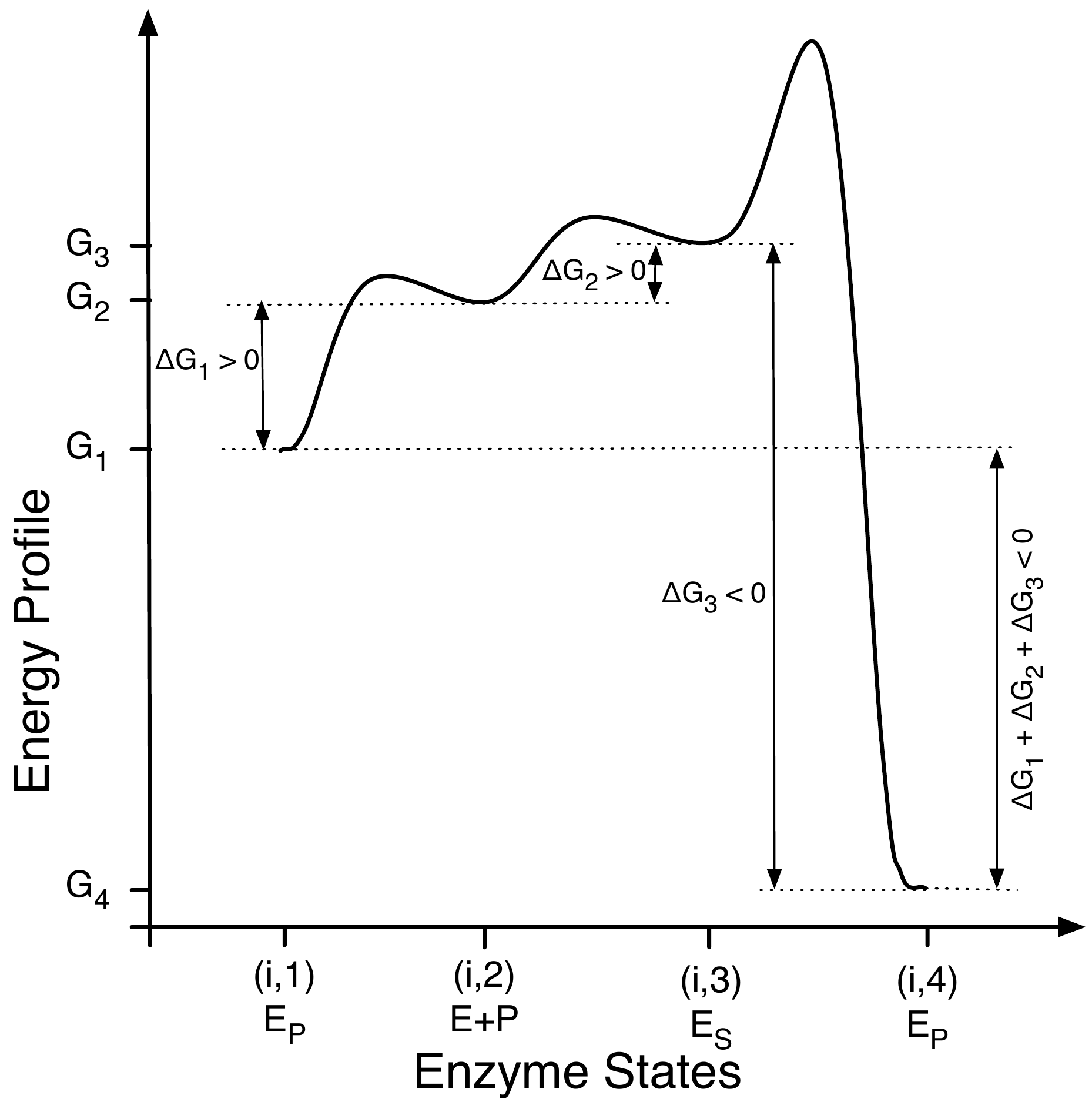}
\caption{A cartoon representation of the Gibbs free energy profile for an enzymatic reaction. The minima correspond to the states the enzyme goes through, while the transition probabilities are determined by the height of the energy barriers. The individual free energy changes between adjacent states can be either positive or negative, but the global free energy change has to be negative in order for the reaction to proceed forward.}
\label{EnergyProfile}
\end{figure}

To better understand the result in Equation (\ref{relrate}) let us analyze the significance of parameters $K_i$ ($i=1,2,3$). From their definition, these parameters are nothing else but the association constants of the following reactions: $E_P \rightleftharpoons E + P$ ($K_1$), $E + S \rightleftharpoons E_S$ ($K_2$), and $E_S \rightleftharpoons E_P$ ($K_3$), respectively. On the other hand, a reaction's association rate ($K_A$) is related to its free energy change ($\Delta G$) by $K_A=\exp (-\Delta G/RT)$, with $R$ the ideal gas constant and $T$ the absolute temperature. These considerations allow us to visualize an enzymatic reaction as a process occurring along an energy profile like the one pictured in Figure \ref{EnergyProfile}. In such scheme, the enzyme states correspond to the local minima of the energy profile, and the transition probabilities are determined by the height of the energy barriers. 

The global free energy change ($\Delta G_T$) of an enzymatic reaction is given by $\Delta G_T = \Delta G_1 + \Delta G_2 + \Delta G_3 = -RT \log (K_1 K_2 K_3)$. Therefore, since the presence of an enzyme does not change $\Delta G_T$:
\[
K_1 K_2 K_3 = \exp \left( - \frac{\Delta G_T}{RT} \right) \equiv \text{constant}.
\]
This restriction further implies that only two of the three $K_i$ constants are independent. Without loss of generality we shall consider that $K_1$ and $K_2$ are determined by the nature of the enzyme catalyzing the reaction, while $K_3 = \exp (-\Delta G/RT) / K_1 K_2$. With this, Equation (\ref{relrate}) can be rewritten as
\begin{equation}
\xi \propto \gamma \frac{\exp (-\Delta G_T/RT)-1}{1+K_1+K_1K_2}.
\label{xi}
\end{equation}
Note that the relaxation rate is a monotonic decreasing function of both $K_1$ and $K_2$. If we further take into consideration that $K_1, K_2 >0$, it follows that the relaxation rate can be increased by making $K_1$ and $K_2$ as close to zero as possible, and therefore by making $\Delta G_1$ and $\Delta G_2$ as positively large as possible. In particular, the maximum value of $\xi$ is attained when $K_1=0$, regardless the value of $K_2$. Thus, in order to increase the value of $\xi$ it is more important to increase the value of $\Delta G_1$ than that of $\Delta G_2$.

On the other hand, we can see from Equation (\ref{xi}) that the relaxation rate is a monotonic decreasing function of $\Delta G_T$. That is, the more energetically favorable the global function is, the more stable the stationary distribution becomes. Similarly, $\xi$ is proportional to $\gamma = -k_3$. Therefore, the system stationary distribution is more strongly stable when the substrate-to-product conversion process is more rapid.

\section{Reaction Velocity and Efficiency of Energy Conversion}

As we have seen, by exploiting the time-scale separation to simplify the reaction scheme, an enzyme can be modeled as a biased one-dimensional random walk with forward and backward transition probabilities $k^+$ and $k^-$, respectively---see Equations (\ref{cmered}) and (\ref{kpkm}). Therefore, by taking into consideration that the stationary probability distribution $P^*$ is constant, the reaction velocity can be defined as the forward minus the backward fluxes:
\begin{equation}
\nu = P^* (k^+ - k^-) = \gamma P^* \frac{\exp (-\Delta G_T/RT)-1}{1+K_1+K_1K_2} .
\label{nu}
\end{equation}

On the other hand, if we consider that a certain amount of energy is consumed during each forward step, and that this energy is waisted during backward steps, the system efficiency in the stationary state can be define as
\begin{equation}
\eta = 1- \frac{k^-}{k^+} = 1 - \frac{1}{\exp (-\Delta G_T/RT)} .
\label{eta}
\end{equation}

Observe that $\nu>0$ is a decreasing function of $\Delta G_T$, and that $\nu=0$ when $\Delta G_T=0$. In other words $\Delta G_T < 0$ in order to have a positive reaction velocity. The efficiency $\eta$ is also a decreasing function of $\Delta G_T$, and $\eta=0$ when $\Delta G_T=0$. That is, both the reaction velocity and its efficiency can be increased by making $\Delta G_T$ more negative.

Note from equations (\ref{xi}) and (\ref{eta}) that $\eta \propto \xi$. Therefore, the discussion regarding the dependence of $\xi$ on $K_1$ and $K_2$ applies as well to $\eta$. In particular, we want to emphasize that the reaction velocity can be increased by making $K_1$ and $K_2$ as close to zero as possible (and thus by making $\Delta G_1$ and $\Delta G_2$ as large as possible). However, varying $K_1$ is more important since the maximum velocity can be achieved by setting $K_1=0$, regardless the value of $K_2$.

Interestingly, the reaction efficiency is independent of $K_1$ and $K_2$. Thus, given that an enzyme does not alter the global free energy change of the reaction it catalyzes, this result implies that a reaction efficiency is the same regardless whether it is catalyzed or not.

\section{Concluding Remarks}

In this work we have investigated the dynamic stability, as well as velocity and efficiency, of an enzymatic reaction with a rate limiting step, at the single molecule level. For this, we followed the ideas in \citep{Esposito:2007uq, Ge:2010kx}, and used a recently developed formalism for performing multiscale thermodynamic analysis on discrete-state, continuos-time, Markovian stochastic processes \citep{SantillanQian2011}. Our results can be summarized as follows:

\begin{enumerate}
\item The dynamic and thermodynamic characteristics associated to the stationary probability distribution are completely determined by the the Gibbs free energy changes of the enzymatic reaction steps: $\Delta G_1$ ($E_P \rightleftharpoons E+ P$), $\Delta G_2$ ($E + S \rightleftharpoons E_S$), and $\Delta G_3$ ($E_S \rightleftharpoons E_P$). Observe that the energies $\Delta G_i$ ($i=1, 2, 3$) are not all independent becase $\sum_{i=1}^3 \Delta G_i = \Delta G_T$, and $\Delta G_T$ (the global free energy change) is not modified by the enzyme. 

\item The stationary probability distribution is stable and the corresponding relaxation rate ($\xi$) is directly proportional to the global reaction velocity ($\nu$).

\item Both $\xi$ and $\nu$ are decreasing functions of $\Delta G_T$, and $\xi, \nu =0$  when $\Delta G_T =0$. Thus, the global reaction accelerates and the stationary probability distribution turns more strongly stable as $\Delta G_T$ is more negative.

\item Both $\xi$ and $\nu$ are increasing functions of $\Delta G_1$ and $\Delta G_2$. The relaxation rate and the reaction velocity achieve their maximum value in the limit $\Delta G_1 \to  \infty$, regardless the value of $\Delta G_2$. Contrarily, $\xi$ and $\nu$ increase as $\Delta G_2$ increases and converge to a value that depends on $\Delta G_1$ as $\Delta G_2 \to \infty$.

\item The efficiency ($\eta$) is a function of $\Delta G_T$, but it is independent of $\Delta G_1$ and $\Delta G_2$. In particular, $\eta$ is a decreasing function of $\Delta G_T$, and $\eta=0$ when $\Delta G_T=0$. That is, the reaction efficiency increases as $\Delta G_T$ becomes more negative. 
\end{enumerate} 

Regarding the novelty of the above results, it is worth pointing out that, while some of them (like the one stating that $\nu$ decreases as $\Delta G_T$ increases) are well known, there are others (like those regarding the relaxation rate $\xi$) which are new to the best of our knowledge. 

On the other hand, when the separation of time scales is not possible, the system dynamic and thermodynamic characteristics will be determined by the rate constants $k^+_i$ and $k^-_i$ ($i=1,2,3$), but the variables $\xi$, $\nu$, and $\eta$, shall not necessarily be given in terms of the equilibrium dissociation constants $K_i = k^-_i / k^+_i$, as in Equations (\ref{xi})-(\ref{eta}). The reason being that the chemical reactions taking place between the synthesis of one product molecule and the next won't necessarily be close to chemical equilibrium. Some of the formerly enumerated results may still be qualitatively true for enzymatic reactions without a rate limiting step, though. However, the study of such reactions is beyond the scope of the present work. In the forthcoming paragraphs we analyze the implications of the results listed above; in the understanding that, unless otherwise stated, these implications may not be exclusive for enzymatic reactions with a rate limiting step.

Although in an enzymatic reaction the global free energy change is predetermined by the substrate, the products, and the nature of their reaction, the actual shape of the energy profile along the reaction coordinate depends on the enzyme structure. Our results confirm that although the efficiency is not affected by the shape of the energy profile, the reaction velocity and the strength of the stationary distribution stability can be highly improved  by properly shaping this profile. This behavior is contrary to that observed in other systems (like thermal engines) in which variation in some parameters makes the velocity and the efficiency change in opposite directions. In other words, in this case no trade-off between efficiency and reaction velocity (and stability strength) need to be looked for, regarding the energy profile. We believe this may by one of the reasons why it has been possible for evolution to drive the structure of enzymes so the corresponding reaction velocity is increased by several orders of magnitude. 

Another feature worth noticing is the fact that the more energetically unfavorable reactions $E_P \rightleftharpoons E+ P$ and $E + S \rightleftharpoons E_S$ are, the faster the global enzymatic reaction is. This behavior can be understood by looking at Figure (\ref{EnergyProfile}). We can see there that large, positive $\Delta G_1$ and $\Delta G_2$ values imply a very negative $\Delta G_3$. This further means that the backward reaction $E_S \leftharpoondown E_P$ is much less probable than the forward reaction $E_S \rightharpoonup E_P$. That is, the strategy to accelerate the global enzymatic reaction seems to be making the rate limiting step almost unidirectional, even though this implies that, in the rapid processes, the backward reactions are more probable than the corresponding forward reactions. Given the prominent role the rate limiting step takes in the above discussion, we believe it does not apply to other types of enzymatic reactions.  

\begin{acknowledgments}
This work was partially supported by Conacyt, M\'{e}xico, Grant: 55228. The author is thankful to the International Centre for Theoretical Physics for its support during the realization of this work.
\end{acknowledgments}

\bibliography{StabVsThermoEnzyme}

\end{document}